\documentstyle[epsfig,preprint,aps]{revtex}           
\begin{document}       
\tighten
           
\def\bea{\begin{eqnarray}}          
\def\eea{\end{eqnarray}}          
\def\beas{\begin{eqnarray*}}          
\def\eeas{\end{eqnarray*}}          
\def\nn{\nonumber}          
\def\ni{\noindent}          
\def\G{\Gamma} 
\def\L{\Lambda}         
\def\d{\delta}          
\def\l{\lambda}          
\def\g{\gamma}          
\def\m{\mu}          
\def\n{\nu}          
\def\s{\sigma}          
\def\tt{\theta}          
\def\b{\beta}          
\def\a{\alpha}          
\def\f{\phi}          
\def\fh{\phi}          
\def\y{\psi}          
\def\z{\zeta}          
\def\p{\pi}          
\def\e{\epsilon}          
\def\ve{\varepsilon}
\def\cf{{\cal F}}  
\def\cg{{\cal G}}  
\def\ch{{\cal H}}
\def\ci{{\cal I}}            
\def\cl{{\cal L}}          
\def\cv{{\cal V}}          
\def\cz{{\cal Z}}          
\def\pl{\partial}          
\def\ov{\over}          
\def\~{\tilde}          
\def\rar{\rightarrow}          
\def\lar{\leftarrow}          
\def\lrar{\leftrightarrow}          
\def\rra{\longrightarrow}          
\def\lla{\longleftarrow}          
\def\8{\infty}          
\newcommand{\fr}{\frac}

\title{Evaluation of a Class of Two-Scale Three-Loop Vacuum Diagrams}

\author{J. -M. Chung\footnote{Email address: jmchung@khu.ac.kr}}
\address{Research Institute for Basic Sciences
and Department of Physics,\\ Kyung Hee University, Seoul 130-701, Korea}
\author{B. K. Chung\footnote{Email address: bkchung@khu.ac.kr}}
\address{Asia Pacific Center for Theoretical Physics, 
Pohang 790-784, Korea\\
and Research Institute for Basic Sciences
and Department of Physics,\\ Kyung Hee University, Seoul 130-701, Korea} 
                
\maketitle              
\draft             
\begin{abstract}           
\indent           
As a generalization of a previous work [Phys. Rev. D. {\bf 59}, 105014
(1999)], 
we compute analytically a class of three-loop vacuum diagrams 
with two {\em arbitrarily} different mass scales. We use a decomposition 
algorithm in which the integrand of the final integral for the third
momentum vector,
say, $k$, becomes independent of the angles of $k$-vector in spherical polar
coordinates. This algorithm proves to be very efficient in obtaining 
exclusively all $\e$-pole terms of the given diagram.   
\end{abstract}               
                   
Vacuum diagrams appear in the evaluation of the effective potential 
as well as in low-energy limits of certain physical
amplitudes or Taylor coefficients of multipoint Green functions. After
the algebra of Dirac gamma matrices and Lorentz indices is performed, 
any vacuum diagram can be  
reduced to some linear combinations of scalar vacuum integrals \cite{av}. 
By the Chetyrkin-Misiak-M\"{u}nz algorithm \cite{cmm}, the computation of 
{\em even} non-vacuum diagrams can be reduced to the problem of the scalar 
vacuum diagrams.    

The authors have computed, using the Avdeev algorithm of
recurrence relations \cite{av}, the single-(mass-)scale three-loop
vacuum diagrams appearing in the three-loop effective 
potential calculation of the massive $\f^4$ theory to the order of
the finite term in the $\e$ expansion \cite{JK1}. In the minimal subtraction 
scheme, knowledge of the {\em finite} terms as well as the pole terms
is needed in the renormalization of the  effective potential \cite{JK1,JMC1,JKP2}.
However, in nonminimal subtraction schemes, the so-called `renormalization conditions' 
are imposed on the effective potential \cite{JMC2,JKP3,JMC3,kt}. In
these schemes, it is sufficient to know only the pole terms of the diagrams.

It is very difficult to apply the Avdeev
algorithm, which was written for single-scale three-loop vacuum
diagrams, 
to the computation of two-scale three-loop diagrams. For the effective
potentials of the $O(N)$ $\f^4$ theory, it is necessary to
compute two-scale three-loop vacuum diagrams \cite{jk}. Note that for the
massive $O(N)$ $\f^4$ theory, the mass-squared ratio of the two scales
is arbitrary whereas for the case of the massless $O(N)$ $\f^4$
theory, the ratio is $1/3$. 

In a previous paper \cite{JMC3}, we computed a class of
two-scale three-loop vacuum diagrams with mass-squared  ratio of $1/3$
by applying the Chetyrkin-Misiak-M\"{u}nz algorithm, which was written
also for a single-scale three-loop vacuum diagrams, to our two-scale 
problems. The purpose of this work is to evaluate analytically three-loop
vacuum diagrams with two {\em arbitrarily} different mass scales by
use of
the Chetyrkin-Misiak-M\"{u}nz algorithm. In this algorithm,
the integrand of the final integral for the third momentum vector,
say, $k$, becomes independent of the angles of $k$-vector in spherical polar
coordinates. This algorithm proves to be very efficient in obtaining 
exclusively all $\e$-pole terms of the given diagram.

The integrals we consider are the following seven nontrivial vacuum
integrals with two mass scales ($m_A$ and $m_B$):
\bea          
J(b)&=&\int_{k}\cg (k^2;m_A^2)\cg (k^2;m_B^2)
=\int_{k}[\ch(k^2;m_A^2,m_B^2)]^2\;,\nn\\      
K(b)&=&\int_{k}{1\ov (k^2+m_A^2)}\cg (k^2;m_A^2)\cg (k^2;m_B^2)\;,\nn\\            
K(c)&=&\int_{k}{1\ov (k^2+m_B^2)}[\ch (k^2;m_A^2,m_B^2)]^2\;,\nn\\  
K(d)&=&\int_{k}{1\ov (k^2+m_A^2)}[\cg (k^2;m_B^2)]^2\;,\nn\\
L(b)&=&\int_{k}{1\ov (k^2+m_A^2)^2}\cg (k^2;m_A^2)\cg (k^2;m_B^2)\;,\nn\\                  
L(c)&=&\int_{k}{1\ov (k^2+m_B^2)^2}[\ch (k^2;m_A^2,m_B^2)]^2\;,\nn\\                  
L(d)&=&\int_{k}{1\ov (k^2+m_A^2)^2}[\cg (k^2;m_B^2)]^2\;,\label{7int}
\eea          
where $\cg (k^2;x)$ $(x=m_A^2, m_B^2)$ and $\ch
(k^2;m_A^2,m_B^2)$
are propagator-type one-loop integrals:
\bea          
\cg (k^2;x)&\equiv&\int_p{1\ov (p^2+x)[(p+k)^2+x]}\;,\nn\\     
\ch (k^2;m_A^2,m_B^2)&\equiv&\int_p{1\ov (p^2+m_A^2)[(p+k)^2+m_B^2]}\;.
~\label{gh}
\eea
Throughout the paper, the
momenta appearing in the formulas are all (Wick-rotated) Euclidean
ones. The abbreviated integration measure is defined as 
\beas
\int_k =\m^{4-D} \int {d^D k\ov (2\p)^D},
\eeas
where $D=4-\e$ is the space-time dimension in the framework of
dimensional regularization and $\m$ is an arbitrary constant with mass dimension.
Notice that the routing of (internal) momenta for $J(b)$ is possible 
in either way given in Eq.~(\ref{7int}).

The functions  $\cg (k^2;x)$ and $\ch(k^2;m_A^2,m_B^2)$ in
Eq.~(\ref{gh}) have simple poles in $\e$:\footnote{Although 
the exact forms of  $G(k^2;x)$ and       
$H(k^2;m_A^2,m_B^2)$ are available \cite{bd}, it is a very difficult
task to carry out the $k$-integrations in Eq.~(\ref{exact}) below exactly.}
\bea          
\cg (k^2;x)&=&{1\ov (4\p)^2}\biggl({x\ov 4\p\m^2}\biggr)^{\!\!\!{-\e/2}}               
\biggl[{2\ov \e}+G(k^2;x)\biggr]\;,\nn \\    
\ch (k^2;m_A^2,m_B^2)
&=&{1\ov (4\p)^2}\biggl({m_A^2\ov 4\p\m^2}\biggr)^{\!\!\!{-\e/2}}               
\biggl[{2\ov \e}+H(k^2;m_A^2,m_B^2)\biggr]\;.\label{dch}              
\eea          
With these splittings, we decompose all of the integrals in
Eq.~(\ref{7int}) as follows:
\bea          
J(b)&=&                
{1\ov (4\p)^4}\biggl({m_A^2\ov 4\p\m^2}\biggr)^{\!\!\!{-\e/2}}               
\biggl({m_B^2\ov 4\p\m^2}\biggr)^{\!\!\!{-\e/2}}               
\int_k G(k^2;m_A^2)G(k^2;m_B^2)\nn\\               
&&+{2W_2\ov (4\p)^2\e}\biggl({m_A^2\ov 4\p\m^2}\biggr)^{\!\!\!{-\e/2}}           
+{2W_1\ov (4\p)^2\e}\biggl({m_B^2\ov 4\p\m^2}\biggr)^{\!\!\!{-\e/2}}          
\;,\nn\\              
K(b)&=&                     
{1\ov (4\p)^4}\biggl({m_A^2\ov 4\p\m^2}\biggr)^{\!\!\!{-\e/2}}               
\biggl({m_B^2\ov 4\p\m^2}\biggr)^{\!\!\!{-\e/2}}               
\int_k{G(k^2;m_A^2)G(k^2;m_B^2)\ov k^2+m_A^2}\nn\\               
&&+{2W_5\ov (4\p)^2\e}\biggl({m_A^2\ov 4\p\m^2}\biggr)^{\!\!\!{-\e/2}}                
+{2W_4\ov (4\p)^2\e}\biggl({m_B^2\ov 4\p\m^2}\biggr)^{\!\!\!{-\e/2}}              
-{4S_1\ov (4\p)^4\e^2}\biggl({m_A^2\ov 4\p\m^2}\biggr)^{\!\!\!{-\e/2}}               
\biggl({m_B^2\ov 4\p\m^2}\biggr)^{\!\!\!{-\e/2}}\;,\nn\\            
L(b)&=&                    
{1\ov (4\p)^4}\biggl({m_A^2\ov 4\p\m^2}\biggr)^{\!\!\!{-\e/2}}               
\biggl({m_B^2\ov 4\p\m^2}\biggr)^{\!\!\!{-\e/2}}               
\int_k{G(k^2;m_A^2)G(k^2;m_B^2)\ov (k^2+m_A^2)^2}\nn\\               
&&+{2W_8\ov (4\p)^2\e}\biggl({m_A^2\ov 4\p\m^2}\biggr)^{\!\!\!{-\e/2}}                
+{2W_6\ov (4\p)^2\e}\biggl({m_B^2\ov 4\p\m^2}           
\biggr)^{\!\!\!{-\e/2}}               
-{4S_3\ov (4\p)^4\e^2}\biggl({m_A^2\ov 4\p\m^2}\biggr)^{\!\!\!{-\e/2}}               
\biggl({m_B^2\ov 4\p\m^2}\biggr)^{\!\!\!{-\e/2}}\;,\nn\\          
J(b)&=&         
{1\ov (4\p)^4}\biggl({m_A^2\ov 4\p\m^2}\biggr)^{\!\!\!{-\e}}             
\int_k \Bigl(H(k^2;m_A^2,m_B^2)\Bigr)^{\!2}+{4W_3\ov (4\p)^2\e}             
\biggl({m_A^2\ov 4\p\m^2}\biggr)^{\!\!\!{-\e/2}}\;,\nn\\           
K(c)&=&                
{1\ov (4\p)^4}\biggl({m_A^2\ov 4\p\m^2}\biggr)^{\!\!\!{-\e}}               
\int_k{[H(k^2;m_A^2,m_B^2)]^2\ov k^2+m_B^2}               
+{4W_5\ov (4\p)^2\e}\biggl({m_A^2\ov 4\p\m^2}\biggr)^{\!\!\!{-\e/2}}               
-{4S_2\ov (4\p)^4\e^2}\biggl({m_A^2\ov 4\p\m^2}\biggr)^{\!\!\!{-\e}}\;,\nn\\            
L(c)&=&          
{1\ov (4\p)^4}\biggl({m_A^2\ov 4\p\m^2}\biggr)^{\!\!\!{-\e}}               
\int_k{[H(k^2;m_A^2,m_B^2)]^2\ov (k^2+m_B^2)^2}               
+{4W_7\ov (4\p)^2\e}\biggl({m_A^2\ov 4\p\m^2}\biggr)^{\!\!\!{-\e/2}}               
-{4S_4\ov (4\p)^4\e^2}\biggl({m_A^2\ov 4\p\m^2}           
\biggr)^{\!\!\!{-\e}}\;,\nn\\          
K(d)&=&          
{1\ov (4\p)^4}\biggl({m_B^2\ov 4\p\m^2}\biggr)^{\!\!\!{-\e}}               
\int_k{[G(k^2;m_B^2)]^2\ov k^2+m_A^2}               
+{4W_5\ov (4\p)^2\e}\biggl({m_B^2\ov 4\p\m^2}\biggr)^{\!\!\!{-\e/2}}               
-{4S_1\ov (4\p)^4\e^2}\biggl({m_B^2\ov 4\p\m^2}           
\biggr)^{\!\!\!{-\e}}\;,\nn\\          
L(d)&=&          
{1\ov (4\p)^4}\biggl({m_B^2\ov 4\p\m^2}\biggr)^{\!\!\!{-\e}}               
\int_k{[G(k^2;m_B^2)]^2\ov (k^2+m_A^2)^2}               
+{4W_8\ov (4\p)^2\e}\biggl({m_B^2\ov 4\p\m^2}\biggr)^{\!\!\!{-\e/2}}               
-{4S_3\ov (4\p)^4\e^2}\biggl({m_B^2\ov 4\p\m^2}           
\biggr)^{\!\!\!{-\e}}\;. \label{exact}                     
\eea          
Here, $S_1$ to $S_4$ and $W_1$ to
$W_8$ are lower-loop integrals \cite{2lp}, whose definitions and values are
relegated to the Appendix. In an abuse of terminology, 
we will call each first term of 
the right-hand sides of Eq.~(\ref{exact}) a `genuine' three-loop 
integral for the corresponding three-loop vacuum diagram.
A remarkable feature of the above
decompositions is that the integrands of genuine three-loop integrals
are, in spheical polar coordinates, independent of the angle variables 
of the $(4-\e)$-dimensional $k$-vector.

A finite volume integration of each $k$ integral in
Eq.~(\ref{exact}) cannot give any $\e$-pole term. Such a pole term can only
arise from integration over large $k^2$ in
Eq.~(\ref{exact}). Therefore, knowing the behaviors of $G(k^2;x)$ 
and $H(k^2;m_A^2,m_B^2)$ at large $q^2$ is enough to find the
UV-divergent
parts of all of the integrals in Eq.~(\ref{7int}). The two-scale
propagator-type one-loop inegral, $\ch(k^2;m_A^2,m_B^2)$, 
can be expanded as an infinite sum of single-scale
propagator-type one-loop integrals:         
\bea          
\int_p{1\ov (p^2+m_A^2)[(p+k)^2+m_B^2]}               
&=&\sum_{\a=0}^\8\biggl(1-{m_B^2\ov m_A^2}\biggr)^\a (m_A^2)^{\a}          
\int_p{1\ov (p^2+m_A^2)[(p+k)^2+m_A^2]^{1+\a}}\;.\label{ep}         
\eea    
The propagator-type one-loop integral with an arbitrary integer power 
of the second propagator was calculated in a closed form \cite{bd}:
\bea
\int_p{1\ov (p^2+m_A^2)[(p+k)^2+m_A^2]^{\a}}&=&      
{(m_A^2)^{1-\a}\ov (4\p)^2}\biggl({m_A^2\ov 4\p\m^2}\biggr)^{\!\!\!{
-\e/2}}{\G(\a-1+\e/2)\ov\G(1+\a)}\nn\\
&&\times\,_3F_2\biggl(1,\a,\a-1+{\e\ov 2};
{1\ov 2}+{\a\ov 2},1+{\a\ov 2};{k^2\ov 4m_A^2}\biggr)\;.\label{3f2}
\eea
In order to see the asymptotic behaviors of  $G(k^2;x)$ 
and $H(k^2;m_A^2,m_B^2)$, we must expand the hypergeometric
function $\,_3F_2$ in Eq.~(\ref{3f2}) in an asymptoic form. 
We simply quote the result \cite{cmm}:
\bea          
\int_p{1\ov (p^2+m_A^2)[(p+k)^2+m_A^2]^{\a}}&=&          
{(m_A^2)^{1-\a}\ov (4\p)^2}
\biggl({m_A^2\ov 4\p\m^2}\biggr)^{\!\!\!{-\e/2}}\nn\\
&&\times \biggl[{2\ov \e}\d_{1\a}     
+\sum_{r=0}^\8\biggl\{\biggl({m_A^2\ov k^2}\biggr)^{\!\!r}          
a^{\rm ren}(\a,r)          
+\biggl({m_A^2\ov k^2}\biggr)^{\!\!r+\e/2}b(\a,r)\biggr\}
\biggr]\;,          \label{k1k2}          
\eea          
where          
\beas        
a^{\rm ren}(\a,r) &=& \left\{ \begin{array}{ll}          
\vspace{0.3cm}\displaystyle{-{2\ov \e}\d_{1\a}    
\d_{0r}}~~~\Bigl(\mbox{when~~$r<1$~~~          
or~~${\a+1\ov 2}\le r<\a$}\Bigr)\;,\nn\\          
\displaystyle{-{2\ov \e}\d_{1\a}\d_{0r}          
+{(-1)^{r-1}(r-1)!\,(\a-r-1)!\, \G(\a-r-1+\e/2)\ov          
(\a-1)!\,(r-1)!\,(\a-2r)!}}\nn\\          
\vspace{0.3cm}~~~~~~~~~~~~~~~~~~~~~~~~~~~~~~~~~~~~~~~~~~~~~~~~~~~~          
\Bigl({\rm when~~}1\le r<{\a+1\ov 2}\Bigr)\;,\nn\\          
\displaystyle{-{2\ov \e}\d_{1\a}\d_{0r}     
+\fr{2 (r-1)!\,(2r-\a-1)!\,           
\G(\a-r-1+\e/2)}{(\a-1)!\,(r-1)!\,(r-\a)!}}
~~~        
\Bigl({\rm when~~}r\ge \a\Bigr)\;,          
\end{array} \right.\nn\\           
b(\a,r)&=&\left\{ \begin{array}{ll}          
\vspace{0.3cm}~~0~~~~~~~          
\Bigl({\rm when~~}r<\a-1\Bigr)\;,\nn\\          
\displaystyle{          
{ (-1)^{r-\a+1} \G(1-r-\e/2)\G(\a-r-\e/2)\G(r+\e/2)\ov          
(\a-1)!\,(r-\a+1)!\,\G(\a+1-2r-\e)}}
~~~\Bigl({\rm when~~}r\ge \a-1\Bigr)\;.          
\end{array} \right.         
\eeas    
Note that $a^{\rm ren}(1,0)$ does contain a simple pole which cancels      
the pole of $b(1,0)$ in the expression $F^{\rm ren}_{\a\b}(k^2)$.    
Only a few lowest terms in the expansion of Eq.~(\ref{k1k2}) affect
the pole parts of the considered three-loop integrals.

The case of $\a=1$ in Eq.~(\ref{k1k2}) gives $\cg(k^2,m_A^2)$.
Thus, the asymptotic behavior of $G(k^2;m_A^2)$ is given by
\bea          
G(k^2;m_A^2)=\sum_{r=0}^\8\biggl[\biggl({m_A^2\ov k^2}\biggr)^{\!\!r}a_r          
+\biggl({m_A^2\ov k^2}\biggr)^{\!\!r+\e/2}b_r\biggr]\;,\label{fays}          
\eea 
with
\beas          
a_0 &=& -{2\ov \e}\;,\nn\\       
a_r &=& {2(2r-2)!\,           
\G(-r+\e/2)\ov (r-1)!}    ~~~(\mbox{when~~$r\ge 1$})\;,  \nn\\                
b_r&=& {(-1)^{r}\G^2(1-r-\e/2)\G(r+\e/2)\ov r!\G(2-2r-\e)}
~~~(\mbox{when~~$r\ge 0$})\;.         
\eeas
From this $G(k^2;m_A^2)$, 
the asymptotic behavior of $G(k^2;m_B^2)$ can be written as          
\bea          
G(k^2;m_B^2)=\sum_{r=0}^\8\biggl[\biggl({m_A^2\ov k^2}\biggr)^{\!\!r}          
\biggl({m_B^2\ov m_A^2}\biggr)^{\!\!r}a_r         
+\biggl({m_A^2\ov k^2}\biggr)^{\!\!r+\e/2}          
\biggl({m_B^2\ov m_A^2}\biggr)^{\!\!r+\e/2}b_r\biggr]\;. \label{gays}         
\eea          
From Eqs.~(\ref{ep}) and (\ref{k1k2}), we obtain the asymptotic behavior of 
$H(k^2;m_A^2,m_B^2)$ in a double series form:         
\bea          
H(k^2;m_A^2,m_B^2)&=&\sum_{\a=0}^\8\sum_{r=0}^\8          
\biggl(1-{m_B^2\ov m_A^2}\biggr)^\a           
\biggl[\biggl({m_A^2\ov k^2}\biggr)^{\!\!r}a^{\rm ren}(1+\a,r)
+\biggl({m_A^2\ov k^2}\biggr)^{\!\!r+\e/2}b(1+\a,r)\biggr]\;.
\label{h}          
\eea

We are now in a position to carry out the $k$ integrations in
Eq.~(\ref{exact}). Let us consider first the following integral:
\beas
\int_k{1\ov (k^2+m_A)^n}f\biggl({m_A^2\ov k^2}\biggr)\;,
\eeas
where 
\beas
f\biggl({m_A^2\ov k^2}\biggr)&=&G(k^2;m_A^2)G(k^2;m_B^2)\nn\\
&=&\sum_{r=0}^\8\sum_{s=0}^\8 \biggl({m_A^2\ov k^2}\biggr)^{\!\!r+s}          
\biggl({m_B^2\ov m_A^2}\biggr)^{\!\!s}\biggl[a_r a_s         
+\biggl({m_A^2\ov k^2}\biggr)^{\!\!\e/2}          
\biggl({m_B^2\ov m_A^2}\biggr)^{\!\!\e/2}a_r b_s\nn\\
&&+\biggl({m_A^2\ov k^2}\biggr)^{\!\!\e/2}b_r a_s
+\biggl({m_A^2\ov k^2}\biggr)^{\!\!\e}          
\biggl({m_B^2\ov m_A^2}\biggr)^{\!\!\e/2}b_r b_s\biggr]\;.
\eeas 
By doing first a trivial angle integration in $(4-\e)$-dimensional spherical 
polar coordinates and then changing the remaining
integration variable $k$, the magnitude of the $(4-\e)$-dimensional
momentum, into $m_A^2/k^2$, one can arrive at 
\beas
\int_k{1\ov (k^2+m_A)^n}f\biggl({m_A^2\ov k^2}\biggr)
&=&{(m_A^2)^{2-n}\ov 4\p^2}\biggl({m_A^2\ov
4\p\m^2}\biggr)^{-\e/2}{1\ov \G(2-\e/2)}\nn\\
&&\times\sum_{l=0}^{\8}
\biggl({-n\atop l}\biggr)\biggl\{\int_0^{m_A^2/\L^2}
+\int_{m_A^2/\L^2}^\8\biggr\}
d\biggl({m_A^2\ov
k^2}\biggl)\biggl({m_A^2\ov k^2}\biggr)^{l-3+n+\e/2}f\biggl({m_A^2\ov k^2}\biggr)\;,
\eeas
where the symbol $({-n\atop l})$ is the expansion coefficient of           
$(1+x)^{-n}=\sum_{l=0}^\8({-n\atop l})x^l$. As was mentioned above, 
the finite-volume $k$ integration does not give rise to any $\e$-pole term.
Thus, within the desired accuracy of $\e$-pole terms, 
we discard the finite-volume integration, the second integral 
$\int_{m_A^2/k^2}^\8 d(m_A^2/\L^2)\cdots$. 
From the result of the trivial integration for a single variable 
$m_A^2/k^2$---trivial in the sense that the integrand is a series sum of
{\em power functions} of $m_A^2/k^2$---it is not difficult to sort out pole terms:      
\bea          
&&{1\ov (4\p)^4}\biggl({m_A^2\ov 4\p\m^2}\biggr)^{\!\!\!{-\e/2}}               
\biggl({m_B^2\ov 4\p\m^2}\biggr)^{\!\!\!{-\e/2}}               
\int_k{G(k^2;m_A^2)G(k^2;m_B^2)\ov (k^2+m_A^2)^n}={(m_A^2)^{2-n}\ov (4\p)^6}          
\biggl({m_A^2\ov 4\p\m^2}\biggr)^{\!\!\!-\e}          
\biggl({m_B^2\ov 4\p\m^2}\biggr)^{\!\!\!-\e/2}\nn\\          
&&~~~~~~~~~~~~~\times           
{2\ov\e\G(2-\e/2)}\sum_{r=0}^{2-n}\,\,\,\sum_{s=0}^{2-n-r}\biggl(          
{-n\atop 2-n-r-s}\biggr)\biggl[a_r a_s          
\biggl({m_B^2\ov m_A^2}\biggr)^{\!\!s}                  
+{1\ov 3} b_r b_s         
\biggl({m_B^2\ov m_A^2}\biggr)^{\!\!s+\e/2}\nn\\          
&&~~~~~~~~~~~~~~~~~~~~~~~          
+{1\ov 2}\biggl\{a_r b_s         
\biggl({m_B^2\ov m_A^2}\biggr)^{\!\!s+\e/2}+          
 b_ra_s\biggl({m_B^2\ov m_A^2}\biggr)^{\!\!s}\biggr\}          
\biggr]+\mbox{[finite terms]}\;.   \label{FG}       
\eea          
From this equation, three integrals corresponding to $n=0,1,2$ are           
evaluated as follows:          
\bea          
&&{1\ov (4\p)^4}\biggl({m_A^2\ov 4\p\m^2}\biggr)^{\!\!\!{-\e/2}}               
\biggl({m_B^2\ov 4\p\m^2}\biggr)^{\!\!\!{-\e/2}}               
\int_k G(k^2;m_A^2)G(k^2;m_B^2)=\Omega_2\biggl[-{16\ov 3\e^3}
\biggl(1-{m_B^2\ov m_A^2}\biggr)^2\nn\\
&&~~~~~~~~~~~~~~
+{1\ov \e^2}\biggl\{\biggl({64\ov 3}-16\g\biggr){m_B^2\ov m_A^2}
+\biggl(-{10\ov 3}
+4\g\biggr)\biggl(1+\biggl({m_B^2\ov m_A^2}\biggr)^2\biggr)
+4\biggl(1-{m_B^2\ov m_A^2}\biggr)^2\ln
\biggl({m_B^2\ov m_A^2}\biggr)\biggr\}\nn\\               
&&~~~~~~~~~~~~~~             
+{1\ov \e}\biggl\{\biggl({80\ov 3}
-32\g+12\g^2+{2\p^2\ov 3}\biggr){m_B^2\ov m_A^2}
+\biggl(-{11\ov 6}+\g-\g^2-{\p^2\ov 6}\biggr)
\biggl(1+\biggl({m_B^2\ov m_A^2}\biggr)^2\biggr)\nn\\
&&~~~~~~~~~~~~~~
+\biggl(4-4\g+(-16+12\g){m_B^2\ov m_A^2}
+(1-2\g)\biggl({m_B^2\ov m_A^2}\biggr)^2\biggr)
\ln\biggl({m_B^2\ov m_A^2}\biggr)\nn\\
&&~~~~~~~~~~~~~~ 
-\biggl(1-{m_B^2\ov m_A^2}\biggr)^2\ln^2
\biggl({m_B^2\ov m_A^2}\biggr)\biggr\}+O(\e^0)           
\biggr]\;,\nn\\            
&&{1\ov (4\p)^4}\biggl({m_A^2\ov 4\p\m^2}\biggr)^{\!\!\!{-\e/2}}               
\biggl({m_B^2\ov 4\p\m^2}\biggr)^{\!\!\!{-\e/2}}               
\int_k{G(k^2;m_A^2)G(k^2;m_B^2)\ov k^2+m_A^2}=\Omega_1\biggl[{1\ov \e^3}\biggl({8\ov 3}
+{16\ov 3}{m_B^2\ov m_A^2}\biggr)\nn\\
&&~~~~~~~~~~~~~~               
+{1\ov \e^2}\biggl\{{16\ov 3}-4\g+(4-4\g){m_B^2\ov m_A^2}
-\biggl(2+4{m_B^2\ov m_A^2}\biggr)\ln\biggl({m_B^2\ov m_A^2}\biggr)
\biggr\}\nn\\               
&&~~~~~~~~~~~~~~             
+{1\ov \e}\biggl\{-2+\g^2+{\p^2\ov 6}
+\biggl(-{8\ov 3}+\g^2+{\p^2\ov 6}\biggr){m_B^2\ov m_A^2}+
\biggl(-7+4\g+2\g{m_B^2\ov m_A^2}\biggr)
\ln\biggl({m_B^2\ov m_A^2}\biggr)\nn\\
&&~~~~~~~~~~~~~~
+\biggl({1\ov 2}+{m_B^2\ov m_A^2}\biggr)\ln^2\biggl({m_B^2\ov m_A^2}\biggr)
\biggr\}+O(\e^0)\biggr]\;,\nn\\           
&&{1\ov (4\p)^4}\biggl({m_A^2\ov 4\p\m^2}\biggr)^{\!\!\!{-\e/2}}               
\biggl({m_B^2\ov 4\p\m^2}\biggr)^{\!\!\!{-\e/2}}               
\int_k{G(k^2;m_A^2)G(k^2;m_B^2)\ov (k^2+m_A^2)^2}=\Omega_0\biggl[{8\ov 3\e^3}               
+{1\ov \e^2}\biggl\{-{4\ov 3}
-2\ln\biggl({m_B^2\ov m_A^2}\biggr) \biggr\}\nn\\   
&&~~~~~~~~~~~~~~~~~~~~~~~~~~~~~~~~~~~~~~~~~             
+{1\ov \e}\biggl\{-{2\ov 3}+\ln\biggl({m_B^2\ov m_A^2}\biggr)
+{1\ov 2}\ln^2\biggl({m_B^2\ov m_A^2}\biggr)\biggr\}               
+O(\e^0)\biggr]\;.   \label{fg}       
\eea          
In the above, the overall multiplying factors are defined as              
\bea               
\Omega_0={1\ov (4\p)^6}\biggl({m_A^2\ov 4\p\m^2}           
\biggr)^{\!\!\!{-3\e/2}}\;,~~              
\Omega_1={m_A^2\ov (4\p)^6}\biggl({m_A^2\ov 4\p\m^2}           
\biggr)^{\!\!\!{-3\e/2}}\;,             
~~              
\Omega_2={m_A^4\ov (4\p)^6}
\biggl({m_A^2\ov 4\p\m^2}\biggr)^{\!\!\!{-3\e/2}}\;, \label{omega}             
\eea              
and $\g$ is the usual Euler constant, $\g=0.5772156649\cdots$.       
             
With Eq.~(\ref{h}), in a manner similar to that leading to
Eq.~(\ref{FG}), we arrive at              
\beas          
{1\ov (4\p)^4}\biggl({m_A^2\ov 4\p\m^2}\biggr)^{\!\!\!{-\e}}          
\int_k{[H(k^2;m_A^2,m_B^2)]^2\ov (k^2+m_B^2)^n}&=&{(m_A^2)^{2-n}\ov (4\p)^6}          
\biggl({m_A^2\ov 4\p\m^2}\biggr)^{\!\!\!{-3\e/2}}{2\ov\e\G(2-\e/2)}\nn\\          
&&\!\!\!\!\!\!\!\!\!\!\!\!\!\!\!\!\!\!\!\!\!\!\!\!\!\!\!\!\!\!\!\!\!\!\!\!\!          
\!\!\!\!\!\!\!\!\!\!\!\!\!\!\!\!\!\!\!\!\!\!\!\!\!\!\!\!\!\!\!\!\!\!\!          
\times\sum_{\a=0}^\8\sum_{\b=0}^\8\biggl(1-{m_B^2\ov m_A^2}\biggr)^{\a+\b}\,\,          
\sum_{r=0}^{2-n}\,\,\,\sum_{s=0}^{2-n-r}          
\biggl({m_B^2\ov m_A^2}\biggr)^{2-n-r-s}\biggl(          
{-n\atop 2-n-r-s}\biggr)\nn\\          
&&\!\!\!\!\!\!\!\!\!\!\!\!\!\!\!\!\!\!\!\!\!\!\!\!\!\!\!\!\!\!\!\!\!\!\!\!\!          
\!\!\!\!\!\!\!\!\!\!\!\!\!\!\!\!\!\!\!\!\!\!\!\!\!\!          
\times\biggl[a^{\rm ren}(1+\a,r)          
a^{\rm ren}(1+\b,s)+{1\ov 3}b(1+\a,r)b(1+\b,s)\nn\\          
&&\!\!\!\!\!\!\!\!\!\!\!\!\!\!\!\!\!\!\!\!\!\!\!\!\!\!\!\!\!\!          
\!\!\!\!\!\!\!\!\!\!\!\!\!\!\!\!\!\!\!\!\!\!\!\!          
+{1\ov 2}\Bigl\{a^{\rm ren}(1+\a,r)b(1+\b,s)+          
b(1+\a,r)a^{\rm ren}(1+\b,s)\Bigr\}\biggr]\nn\\          
&&\!\!\!\!\!\!\!\!\!\!\!\!\!\!\!\!\!\!\!\!\!\!\!\!\!\!\!\!\!\!          
\!\!\!\!\!\!\!\!\!\!\!\!\!\!\!\!\!         
+\mbox{[finite terms]}\;,        
\eeas               
from which the three integrals for $n=0,1,2$ are calculated as follows:          
\bea             
&&{1\ov (4\p)^4}\biggl({m_A^2\ov 4\p\m^2}\biggr)^{\!\!\!{-\e}}             
\int_k \Bigl(H(k^2;m_A^2,m_B^2)\Bigr)^{\!2}=\Omega_2\biggl[
{8\ov 3\e^3}\biggl(1-{m_B^2\ov m_A^2}\biggr)^2+{1\ov \e^2}               
\biggl\{{16\ov 3}{m_B^2\ov m_A^2}\nn\\
&&~~~~~~~~~~~~~~~~~~
+\biggl({14\ov 3}-4\g\biggr)
\biggl(1+\biggl({m_B^2\ov m_A^2}\biggr)^2\biggr)
-4\biggl({m_B^2\ov m_A^2}\biggr)^2\ln{m_B^2\ov m_A^2}\biggr\}\nn\\               
&&~~~~~~~~~~~~~~~~~~              
+{1\ov \e}\biggl\{\biggl({44\ov 3}-16\g+4\g^2\biggr){m_B^2\ov m_A^2}
+\biggl({25\ov 6}-7\g+3\g^2+{\p^2\ov 6}\biggr)
\biggl(1+\biggl({m_B^2\ov m_A^2}\biggr)^2\biggr) \nn\\
&&~~~~~~~~~~~~~~~~~~
+\biggl(
(-8+4\g){m_B^2\ov m_A^2}+(-7+6\g)\biggl({m_B^2\ov m_A^2}\biggr)^2
\biggr)\ln\biggl({m_B^2\ov m_A^2}\biggr)\nn\\
&&~~~~~~~~~~~~~~~~~~
+3\biggl({m_B^2\ov m_A^2}\biggr)^2\ln^2\biggl({m_B^2\ov m_A^2}\biggr)
\biggr\}+O(\e^0)\biggr]\;,\nn\\          
&&{1\ov (4\p)^4}\biggl({m_A^2\ov 4\p\m^2}\biggr)^{\!\!\!{-\e}}               
\int_k{[H(k^2;m_A^2,m_B^2)]^2\ov k^2+m_B^2}=\Omega_1\biggl[{1\ov \e^3} 
\biggl({16\ov 3}+{8\ov 3}{m_B^2\ov m_A^2}\biggr)             
+{1\ov \e^2}\biggl\{4-4\g+\biggl({16\ov 3}-4\g\biggr){m_B^2\ov m_A^2}\nn\\
&&~~~~~~~~~~~~~~~~~~
-4{m_B^2\ov m_A^2}\ln\biggl({m_B^2\ov m_A^2}\biggr)\biggr\}                        
+{1\ov \e}\biggl\{-{8\ov 3}+\g^2+{\p^2\ov 6}
+\biggl(-2+\g^2+{\p^2\ov 6}\biggr){m_B^2\ov m_A^2}\nn\\
&&~~~~~~~~~~~~~~~~~~
+2\g{m_B^2\ov m_A^2}\ln\biggl({m_B^2\ov m_A^2}\biggr)
+{m_B^2\ov m_A^2}\ln^2\biggl({m_B^2\ov m_A^2}\biggr)
\biggr\}+O(\e^0)\biggr]\;,\nn\\            
&&{1\ov (4\p)^4}\biggl({m_A^2\ov 4\p\m^2}\biggr)^{\!\!\!{-\e}}               
\int_k{[H(k^2;m_A^2,m_B^2)]^2\ov (k^2+m_B^2)^2}=\Omega_0\biggl[{8\ov 3\e^3}               
-{4\ov 3\e^2}-{2\ov 3\e}+O(\e^0)\biggr]\;.\label{hh}          
\eea       

Finally, using the asymptotic form of $G(k^2;m_B^2)$, Eq.~(\ref{gays}), we obtain         
\bea          
&&{1\ov (4\p)^4}\biggl({m_B^2\ov 4\p\m^2}\biggr)^{\!\!\!{-\e}}          
\int_k{[G(k^2;m_B^2)]^2\ov (k^2+m_A^2)^n}={(m_A^2)^{2-n}\ov (4\p)^6}          
\biggl({m_A^2\ov 4\p\m^2}\biggr)^{\!\!\!-\e/2}          
\biggl({m_B^2\ov 4\p\m^2}\biggr)^{\!\!\!-\e}          
{2\ov\e\G(2-\e/2)}\nn\\          
&&~~~~~~~~~~~~~        
\times\sum_{r=0}^{2-n}\,\,\,\sum_{s=0}^{2-n-r}\biggl(          
{-n\atop 2-n-r-s}\biggr)\biggl[a_ra_s          
\biggl({m_B^2\ov m_A^2}\biggr)^{\!\!r+s}
+{1\ov 3} b_r b_s         
\biggl({m_B^2\ov m_A^2}\biggr)^{\!\!r+s+\e}\nn\\          
&&~~~~~~~~~~~~~~~~~~~~~~~          
+{1\ov 2}\biggl\{a_r  b_s         
+ b_r a_s\biggr\}          
\biggl({m_B^2\ov m_A^2}\biggr)^{\!\!r+s+\e/2}\biggr]      
+\mbox{[finite terms]}\;.\nn          
\eea          
After substitutions of the appropriate values          
for $a_r$ and $b_r$, we end up with         
\bea             
&&{1\ov (4\p)^4}\biggl({m_B^2\ov 4\p\m^2}\biggr)^{\!\!\!{-\e}}               
\int_k{[G(k^2;m_B^2)]^2\ov k^2+m_A^2}=\Omega_1\biggl[{1\ov e^3}\biggl(
-{8\ov 3}+{32\ov 3}{m_B^2\ov m_A^2}\biggr)               
+{1\ov \e^2}\biggl\{{4\ov 3}+(8-8\g){m_B^2\ov m_A^2}\nn\\
&&~~~~~~~~~~~~~~~~~~  
+\biggl(4-16{m_B^2\ov m_A^2}\biggr)\ln\biggl({m_B^2\ov m_A^2}\biggr)
\biggr\}+{1\ov \e}\biggl\{{2\ov 3}
+\biggl(-{16\ov 3}+2\g^2+{\p^2\ov 3}\biggr){m_B^2\ov m_A^2}\nn\\
&&~~~~~~~~~~~~~~~~~~
+\biggl(-2+(-12+12\g){m_B^2\ov m_A^2}\biggr)
\ln\biggl({m_B^2\ov m_A^2}\biggr)+\biggl(-3+12{m_B^2\ov m_A^2}\biggr)
\ln^2\biggl({m_B^2\ov m_A^2}\biggr)\biggr\}+O(\e^0)\biggr]\;,\nn\\          
&&{1\ov (4\p)^4}\biggl({m_B^2\ov 4\p\m^2}\biggr)^{\!\!\!{-\e}}          
\int_k{[G(k^2;m_B^2)]^2\ov (k^2+m_A^2)^2}=\Omega_0\biggl[{8\ov 3\e^3}               
+{1\ov \e^2}\biggl\{-{4\ov 3}
-4\ln\biggl({m_B^2\ov m_A^2}\biggr) \biggr\}\nn\\            
&&~~~~~~~~~~~~~~~~~~~~~~~~~~~~~~~~~~~~~~~~~~~              
+{1\ov \e}\biggl\{-{2\ov 3}+2\ln\biggl({m_B^2\ov m_A^2}\biggr)
+3\ln^2\biggl({m_B^2\ov m_A^2}\biggr)\biggr\}+O(\e^0)\biggr]\;.\label{gg}               
\eea

In summary, we tabulate a class of three-loop vacuum integrals given 
in Eq.~(\ref{7int}) by substituting the results for the `genuine' 
three-loop integrals scattered in Eqs.~(\ref{fg}), (\ref{hh}), 
and (\ref{gg}) and for
the `factorized' lower-loop integrals in Eq.~(\ref{www}) into 
Eq.~(\ref{exact}). Since, in Eq.~(\ref{exact}),  
one-loop integrals ($S_1$ to $S_4$) are multiplied by
$1/\e^2$  and  two-loop integrals ($W_1$ to $W_8$) by $1/\e$, it is
sufficient to know
their values to the order of  $\e^1$ and $\e^0$, respectively.
The tabulation reads as follows:    
\bea             
J(b)&=&\Omega_2               
\biggl[{1\ov \e^3}\biggl\{{32\ov 9}+{32\ov 3}{m_B^2\ov m_A^2}
-{16\ov 3}\biggl({m_B^2\ov m_A^2}\biggr)^2\biggr\}
+{1\ov \e^2}\biggl\{\biggl({64\ov 3}-16\g\biggr){m_B^2\ov m_A^2}\nn\\ 
&&+\biggl({14\ov 3}-4\g\biggr)\biggl(1+\biggl({m_B^2\ov m_A^2}\biggr)^2\biggr)
-\biggl(8+4{m_B^2\ov m_A^2}\biggr){m_B^2\ov m_A^2}
\ln\biggl({m_B^2\ov m_A^2}\biggr)\biggr\}\nn\\    
&&+{1\ov \e}\biggl\{\biggl({80\ov 3}-32\g+12\g^2+{2\p^2\ov 3}\biggr)
{m_B^2\ov m_A^2}+\biggl({25\ov 6}-7\g+3\g^2+{\p^2\ov 6}\biggr)
\biggl(1+\biggl({m_B^2\ov m_A^2}\biggr)^2\biggr)\nn\\
&&+\biggl((-16+12\g){m_B^2\ov m_A^2}
+(-7+6\g)\biggl({m_B^2\ov m_A^2}\biggr)^2\biggr)
\ln\biggl({m_B^2\ov m_A^2}\biggr)
+\biggl(2+3{m_B^2\ov m_A^2}\biggr){m_B^2\ov m_A^2} 
\ln^2\biggl({m_B^2\ov m_A^2}\biggr)\biggr\}           
\biggr]\;,\nn\\                 
K(b)&=&\Omega_1\biggl[-{1\ov \e^3}\biggl({16\ov 3}+{8\ov 3}
{m_B^2\ov m_A^2}\biggr)+{1\ov \e^2}\biggl\{-{44\ov 3}+8\g
+\biggl(-8+4\g\biggr){m_B^2\ov m_A^2}
+4{m_B^2\ov m_A^2}\ln\biggl({m_B^2\ov m_A^2}\biggr)\biggr\}\nn\\      
&&+{1\ov \e}\biggl\{-28+22\g-6\g^2-{\p^2\ov 3}
+4\sqrt{3}\,{\rm Ls}_2\Bigl({\p\ov 3}\Bigr)
+\biggl(-{50\ov 3}+12\g-3\g^2-{\p^2\ov 6}\biggr){m_B^2\ov m_A^2}\nn\\
&&+4\biggl(4{m_B^2\ov m_A^2}-1\biggr)^{1/2}{\rm Ls}_2(\tt)
+(12-6\g){m_B^2\ov m_A^2}\ln\biggl({m_B^2\ov m_A^2}\biggr)
+\biggl(1-3{m_B^2\ov m_A^2}\biggr)
\ln^2\biggl({m_B^2\ov m_A^2}\biggr)\biggr\}\biggr]\;,\nn\\                 
K(c)&=&\Omega_1\biggl[-{1\ov \e^3}\biggl({8\ov 3}
+{16\ov 3}{m_B^2\ov m_A^2}\biggr)
+{1\ov \e^2}\biggl\{-8+4\g+\biggl(-{44\ov 3}+8\g\biggr)
{m_B^2\ov m_A^2}
+8{m_B^2\ov m_A^2}\ln\biggl({m_B^2\ov m_A^2}\biggr)\biggr\}\nn\\               
&&+{1\ov \e}\biggl\{-{50\ov 3}+12\g-3\g^2-{\p^2\ov 6}
+\biggl(-28+22\g-6\g^2-{\p^2\ov 3}\biggr){m_B^2\ov m_A^2}\nn\\
&&+8\biggl(4{m_B^2\ov m_A^2}-1\biggr)^{1/2}{\rm Ls}_2(\tt)
+(22-12\g){m_B^2\ov m_A^2}\ln\biggl({m_B^2\ov m_A^2}\biggr)
+\biggl(2-6{m_B^2\ov m_A^2}\biggr)\ln^2\biggl({m_B^2\ov m_A^2}\biggr)
\biggr\}\biggr]\;,\nn\\                 
K(d)&=&\Omega_1\biggl[-{1\ov \e^3}\biggl({8\ov 3}
+{16\ov 3}{m_B^2\ov m_A^2}\biggr)
+{1\ov \e^2}\biggl\{-{20\ov 3}+4\g+\biggl(-16+8\g\biggr)
{m_B^2\ov m_A^2}
+8{m_B^2\ov m_A^2}\ln\biggl({m_B^2\ov m_A^2}\biggr)\biggr\}\nn\\               
&&+{1\ov \e}\biggl\{-{34\ov 3}+10\g-3\g^2-{\p^2\ov 6}
+\biggl({100\ov 3}+24\g-6\g^2-{\p^2\ov 3}\biggr){m_B^2\ov m_A^2}\nn\\
&&+8\biggl(4{m_B^2\ov m_A^2}-1\biggr)^{1/2}{\rm Ls}_2(\tt)
+(24-12\g){m_B^2\ov m_A^2}\ln\biggl({m_B^2\ov m_A^2}\biggr)
+\biggl(2-6{m_B^2\ov m_A^2}\biggr)\ln^2\biggl({m_B^2\ov m_A^2}\biggr)
\biggr\}\biggr]\;,\nn\\                 
L(b)&=&\Omega_0\biggl[{8\ov 3\e^3}+{1\ov \e^2}\biggl\{{8\ov 3}-4\g\biggr\}                 
+{1\ov \e}\biggl\{{4\ov 3}-4\g+3\g^2+{\p^2\ov 6}
-{4\ov \sqrt{3}}{\rm Ls}_2\Bigl({\p\ov 3}\Bigr)\nn\\
&&+\biggl(8\biggl(4{m_B^2\ov m_A^2}-1\biggr)^{-1/2}\,{m_B^2\ov m_A^2}
-4\biggl(4{m_B^2\ov m_A^2}-1\biggr)^{1/2}\biggr){\rm Ls}_2(\tt)
+4\ln|2\sin(\tt/2)|\nn\\
&&+2\ln\biggl({m_B^2\ov m_A^2}\biggr)
-\ln^2\biggl({m_B^2\ov m_A^2}\biggr)\biggr\}\biggr]\;,\nn\\                 
L(c)&=&\Omega_0               
\biggl[{8\ov 3\e^3}+{1\ov \e^2}\biggl\{{8\ov 3}-4\g
-4\ln\biggl({m_B^2\ov m_A^2}\biggr)\biggr\}\nn\\            
&&+{1\ov \e}\biggl\{{4\ov 3 }-4\g+3\g^2+{\p^2\ov 6}
-8\biggl(4{m_B^2\ov m_A^2}-1\biggr)^{-1/2}{\rm Ls}_2(\tt)\nn\\
&&-4\biggl({m_B^2\ov m_A^2}\biggr)^{-1}\ln|2\sin(\tt/2)|
+\biggl(-4+6\g-2\biggl({m_B^2\ov m_A^2}\biggr)^{-1}\biggr)
\ln\biggl({m_B^2\ov m_A^2}\biggr)
+3\ln^2\biggl({m_B^2\ov m_A^2}\biggr)\biggr\}\biggr]\;,\nn\\                 
L(d)&=&\Omega_0\biggl[{8\ov 3\e^3}+{1\ov \e^2}\biggl\{{8\ov 3}-4\g\biggr\}                 
+{1\ov \e}\biggl\{{4\ov 3}-4\g+3\g^2+{\p^2\ov 6}\nn\\
&&+\biggl(16\biggl(4{m_B^2\ov m_A^2}-1\biggr)^{-1/2}\,{m_B^2\ov m_A^2}
-8\biggl(4{m_B^2\ov m_A^2}-1\biggr)^{1/2}\biggr){\rm Ls}_2(\tt)
+8\ln|2\sin(\tt/2)|\nn\\
&&+4\ln\biggl({m_B^2\ov m_A^2}\biggr)-2\ln^2\biggl({m_B^2\ov m_A^2}\biggr)
\biggr\}\biggr]\;,  \label{sum}          
\eea                 
where the overall multiplying factors $\Omega_0$, $\Omega_1$, and
$\Omega_2$ are defined in Eq.~(\ref{omega}) and ${\rm Ls}_2(\tt)$ is 
the log-sine integral appearing in Eq.~(\ref{ls2}).

\section*{Acknowledgment}               
This work was supported by the Brain Korea 21 Project in 2001.

\appendix
\section*{Lower-loop integrals}
We give the definitions of lower-loop integrals $S_1$ to  $S_4$ and $W_1$
to $W_8$ appearing in Eq.~(\ref{exact}) and their
values:
\bea                 
S_1&\equiv&\int_k{1\ov k^2+m_A^2}                 
={m_A^2\ov (4\p)^2}\biggl({m_A^2\ov 4\p\m^2}                 
\biggr)^{\!\!\!-\e/2}\G\biggl({\e\ov 2}-1\biggr)\;,\nn\\                 
S_2&\equiv&\int_k{1\ov k^2+m_B^2}                 
=\biggl({m_B^2\ov m_A^2}\biggr)^{1-\e/2}S_1\;,\nn\\                
S_3&\equiv&
\int_k{1\ov (k^2+m_A^2)^2}={1\ov (4\p)^2}\biggl({m_A^2\ov 4\p\m^2}                 
\biggr)^{\!\!\!-\e/2}\G\biggl({\e\ov 2}\biggr)\;,\nn\\                 
S_4&\equiv&\int_k{1\ov (k^2+m_B^2)^2}
=\biggl({m_B^2\ov m_A^2}\biggr)^{-\e/2}S_3\;,\nn\\                           
W_1&\equiv&\int_{kp}{1\ov (p^2+m_A^2)[(p+k)^2+m_A^2]}  
=S_1^2\;,\nn\\               
W_2&\equiv&\int_{kp}{1\ov (p^2+m_B^2)[(p+k)^2+m_B^2]}
=S_2^2\;\nn\\
W_3&\equiv&\int_{kp}{1\ov (p^2+m_A^2)[(p+k)^2+m_B^2]}
=S_1 S_2\;,\nn\\
W_4&\equiv&\int_{kp}{1\ov (k^2+m_A^2)                 
(p^2+m_A^2)[(p+k)^2+m_A^2]}\nn\\                
&=&{m_A^2\ov (4\p)^4}\biggl({m_A^2\ov 4\p\m^2}\biggr)^{\!\!\!-\e}                 
{\G^2(1+\e/2)\ov (1-\e)(1-\e/2)}\biggl[-{6\ov \e^2}
+2\sqrt{3}{\rm Ls}_2\Bigl({\p\ov 3}\Bigr)+O(\e)           
\biggr]\;,\nn\\                 
W_5&\equiv&               
\int_{kp}{1\ov (k^2+m_A^2)(p^2+m_B^2)[(p+k)^2+m_B^2]}\nn\\                 
&=&{m_A^2\ov (4\p)^4}\biggl({m_A^2\ov 4\p\m^2}\biggr)^{\!\!\!{-\e}}                 
{\G^2(1+\e/2)\ov (1-\e)(1-\e/2)}\biggl[-{2\ov \e^2}\biggl\{2
\biggl({m_B^2\ov m_A^2}\biggr)^{-\e/2}\nn\\
&&+\biggl(2{m_B^2\ov m_A^2}-1\biggr)\biggl({m_B^2\ov m_A^2}\biggr)^{-\e}
\biggr\}+\biggl(4{m_B^2\ov m_A^2}-1\biggr)^{(1-\e)/2}
\biggl\{2{\rm Ls}_2(\tt)+O(\e)\biggr\}\biggr]\;,\nn\\                
W_6&\equiv&               
\int_{kp}{1\ov (k^2+m_A^2)^2(p^2+m_A^2)[(p+k)^2+m_A^2]}\nn\\                 
&=&{1\ov (4\p)^4}\biggl({m_A^2\ov 4\p\m^2}\biggr)^{\!\!\!-\e}
{\G^2(1+\e/2)\ov (1-\e/2)}\biggl[{2\ov \e^2}
-{2\ov \sqrt{3}}{\rm Ls}_2\Bigl({\p\ov 3}\Bigr)+O(\e)\biggr]\;,\nn\\  
W_7&\equiv&\int_{kp}{1\ov (k^2+m_B^2)^2(p^2+m_B^2)[(k+p)^2+m_A^2]}\nn\\                 
&=&{1\ov (4\p)^4}\biggl({m_A^2\ov 4\p\m^2}                 
\biggr)^{\!\!\!-\e}{\G^2(1+\e/2)\ov (1-\e)(1-\e/2)}
\biggl[{2\ov \e^2}\biggl({m_B^2\ov m_A^2}\biggr)^{-\e}
+{1\ov \e}\biggl\{-\biggl({m_B^2\ov m_A^2}\biggr)^{-1-\e/2}\nn\\
&&-\biggl(2{m_B^2\ov m_A^2}-1\biggr)\biggl({m_B^2\ov m_A^2}\biggr)^{-1-\e}
\biggr\}-2\biggl(4{m_B^2\ov m_A^2}-1\biggr)^{-(1+\e)/2}\,{\rm Ls}_2(\tt)\nn\\
&&-\biggl(4{m_B^2\ov m_A^2}-1\biggr)^{-\e/2}
\biggl({m_B^2\ov m_A^2}\biggr)^{-1}
\ln|2\sin(\tt/2)|+O(\e)\biggr]\;,\nn\\                 
W_8&\equiv&               
\int_{kp}{1\ov (k^2+m_A^2)^2(p^2+m_B^2)[(k+p)^2+m_B^2]}\nn\\                 
&=&{1\ov (4\p)^4}\biggl({m_A^2\ov 4\p\m^2}                 
\biggr)^{\!\!\!-\e}{\G^2(1+\e/2)\ov (1-\e)(1-\e/2)}
\biggl[{2\ov \e^2}\biggl\{
2\biggl({m_B^2\ov m_A^2}\biggr)^{-\e/2}
-\biggl({m_B^2\ov m_A^2}\biggr)^{-\e}\biggr\}\nn\\
&&-{2\ov \e}\biggl({m_B^2\ov m_A^2}\biggr)^{-\e/2}
-2\biggl(4{m_B^2\ov m_A^2}-1\biggr)^{(1-\e)/2}{\rm Ls}_2(\tt)
+4\biggl(4{m_B^2\ov m_A^2}-1\biggr)^{-(1+\e)/2}\,{m_B^2\ov m_A^2}
{\rm Ls}_2(\tt)\nn\\
&&+2\biggl(4{m_B^2\ov m_A^2}-1\biggr)^{-\e/2}\ln|2\sin(\tt/2)|
+O(\e)\biggr]\;,\label{www}               
\eea                 
where
\bea 
{\rm Ls}_2(\tt) \equiv -\int_0^\tt d\tt'\ln|2\sin(\tt'/2)|\;, ~~~~
\tt=\cos^{-1}\biggl(1-{1\ov 2}{m_A^2\ov m_B^2}\biggr)\;. \label{ls2}
\eea

 
\end{document}